\documentclass[aps,twocolumn,amssymb,prl,showpacs,10pt]{revtex4-2}

\usepackage[utf8]{inputenc}
\usepackage{tabularx}
\usepackage{multirow}
\usepackage{url}
\usepackage{amsmath}
\usepackage{amssymb}	
\usepackage{gensymb}
\usepackage{graphicx}
\usepackage{mathrsfs}
\usepackage{color}
\usepackage[normalem]{ulem}
\usepackage{float}
\usepackage{lmodern}
\usepackage{soul}
\usepackage[breaklinks,colorlinks,urlcolor=blue,citecolor=blue]{hyperref}

\newcommand{\be}{\begin{equation}}
\newcommand{\ee}{\end{equation}}
\newcommand{\figg}[1]{Fig.~\ref{fig:#1}}
\newcommand{\ex}[1]{10^{-#1}}

\newcommand{\comp}{c/\omega_{\mathrm{p}}}
\newcommand{\omp}{\omega_{\mathrm{p}}}
\def\ompt{\omega_{\rm p}t}
\newcommand{\xsh}{x_{\rm sh}}
\newcommand{\gbx}{\gamma\beta_x}
\newcommand{\gby}{\gamma\beta_y}

\newcommand{\araa}{ARA\&A}
\newcommand{\apjl}{ApJ}
\newcommand{\mnras}{MNRAS}
\newcommand{\aapr}{A\&A~Rev.}
\newcommand{\physrep}{Phys.~Rep.}
\newcommand{\aap}{A\&A}

\newcommand{\ls}[1]{{#1}}

\begin{document}

\title{Coherent Electromagnetic Emission from Relativistic Magnetized 
Shocks}

\author{Lorenzo Sironi}
\email{lsironi@astro.columbia.edu}
\affiliation{Department of Astronomy and Columbia Astrophysics Laboratory, Columbia University, New York, NY 10027, USA}
\author{Illya Plotnikov}
\affiliation{IRAP, Universit\'e de Toulouse III - Paul Sabatier, OMP, Toulouse, France}
\author{Joonas N\"attil\"a}
\affiliation{Physics Department and Columbia Astrophysics Laboratory, Columbia University, 538 West 120th Street, New York, NY 10027, USA}
\affiliation{Center for Computational Astrophysics, Flatiron Institute, 162 Fifth Avenue, New York, NY 10010, USA}
\author{Andrei M. Beloborodov}
\affiliation{Physics Department and Columbia Astrophysics Laboratory, Columbia University, 538 West 120th Street, New York, NY 10027, USA}
\affiliation{Max Planck Institute for Astrophysics, Karl-Schwarzschild-Str. 1, D-85741, Garching, Germany}

\date{\today}


\begin{abstract}
\ls{Relativistic magnetized shocks are a natural source of coherent emission, offering a plausible radiative mechanism for Fast Radio Bursts (FRBs). We present first-principles 3D simulations that provide essential information for the FRB models based on shocks: the emission efficiency, spectrum, and polarization.  The simulated shock propagates in an $e^\pm$ plasma with magnetization $\sigma>1$. The measured fraction of shock energy converted to coherent radiation is $\simeq \ex{3} \, \sigma^{-1}$, and the energy-carrying wavenumber of the wave spectrum is  $\simeq 4 \,\omega_{\rm c}/c$, where $\omega_{\rm c}$ is the upstream gyrofrequency.
The ratio of the O-mode and X-mode energy fluxes emitted by the shock is $\simeq 0.4\,\sigma^{-1}$.
{The dominance of the X-mode  at $\sigma\gg 1$ is particularly strong, approaching 100\% in the spectral band around $2\,\omega_c$.}
We also provide a detailed description of the emission mechanism for both X- and O-modes. 
}
\end{abstract}

\maketitle


The discovery of Fast Radio Bursts \citep[FRBs;][]{petroff_19,cordes_19,platts_19}
 has revived the interest in astrophysical sources of coherent emission \citep{lyubarsky_21}. 
 FRBs are bright ($\sim 1\,$Jy) pulses of millisecond duration in the GHz band, and their extreme brightness temperatures require a coherent emission mechanism \citep[][]{katz_16}. 
Magnetars are commonly invoked as FRB progenitors, a hypothesis recently supported by the detection of FRBs from a Galactic magnetar \citep{scholz_20,bochenek_20}. 
Magnetar flares are capable of driving explosions into the magnetar wind \citep{belo_17,lorenzometzger,belo_20,yuan_20}, resembling shocks in the solar wind launched by solar flares. In contrast to the solar activity, the winds
and explosions from magnetars are ultra-relativistic.
 
Shocks in magnetar winds are strongly magnetized \citep{belo_20}, with magnetization $\sigma\sim 10-100$ ($\sigma$ is the ratio of upstream Poynting flux to kinetic energy flux). 
Relativistic magnetized shocks are a natural source of coherent emission, via the so-called ``synchrotron maser instability'' \citep{alsop_arons_88,hoshino_91}, which generates a train of ``precursor waves'' propagating ahead of the shock%
\footnote{The generation of coherent emission is not directly due to wave amplification via a maser process, but still we shall refer to this as the ``synchrotron maser,'' because this term is widely used in the literature.}. 
The fundamental properties of the precursor waves 
can be quantified with kinetic particle-in-cell (PIC) simulations. 
%

A stringent constraint on any FRB emission mechanism is imposed by the high degree of polarization 
observed in some FRBs \citep[][]{michilli_18}. 
The synchrotron maser generates waves with the X-mode linear polarization (fluctuating electric field perpendicular to the pre-shock magnetic field).
The shock is, however, also able to generate O-mode waves (electric field parallel to the pre-shock magnetic field),
and only 3D simulations can provide a realistic picture of the polarized shock emission. 
In this Letter, we present a suite of 3D PIC simulations of $\sigma\gtrsim 1$ relativistic electron-positron shocks, 
extending earlier 1D and 2D studies
\citep{langdon_88,gallant_92, sironi_spitkovsky_09, iwamoto_17, iwamoto_18,plotnikov_18,plotnikov2019,babul_sironi_20}.
We quantify the efficiency, spectrum and polarization of precursor waves, and provide a detailed description of the X-mode and O-mode emission mechanisms.

\textit{Simulation setup.---}
We use the electromagnetic PIC code TRISTAN-MP \citep{spitkovsky_05} to perform 3D shock simulations in the post-shock frame.
The upstream flow is a cold pair plasma drifting in the $-\hat{x}$ direction with bulk Lorentz factor $\gamma_0=3$ (\ls{selected runs with} $\gamma_0=10$ lead to similar conclusions). 
The shock is launched as the incoming flow reflects off a wall at $x=0$ and propagates along $+\hat{x}$. 

The pre-shock plasma with density $n_0$ carries a frozen-in magnetic field $\boldsymbol{B}_0=B_0\,\hat{z}$ and its motional electric field $-E_0\,\hat{y}$, where $E_0=\beta_0B_0$ and $\beta_0=\sqrt{1-1/\gamma_0^2}$. 
All these quantities are defined in the simulation frame.
The field strength is parameterized via the magnetization $\sigma = B_{0}^2 / 4 \pi \gamma_0 n_0 m c^2 = \omega_{\rm c}^2 /  \omega_{\rm p}^2$,
which we vary in the range $0.6\leq \sigma\leq 10$. 
Here, $\omega_{\rm c}=eB_0/\gamma_0 m c$ is the gyrofrequency, and $\omp=(4 \pi n_0 e^2/\gamma_0 m)^{1/2}$  the plasma frequency. 
{Our reference simulations employ 3 particles per species per cell and a spatial resolution of $\comp=25$ cells (see Suppl.~Mat.). We evolve our simulations for several thousands of $\omp^{-1}$, when both X-mode and  O-mode emissions reach a steady state.}

\begin{figure}
\centering
    \includegraphics[width=\columnwidth]{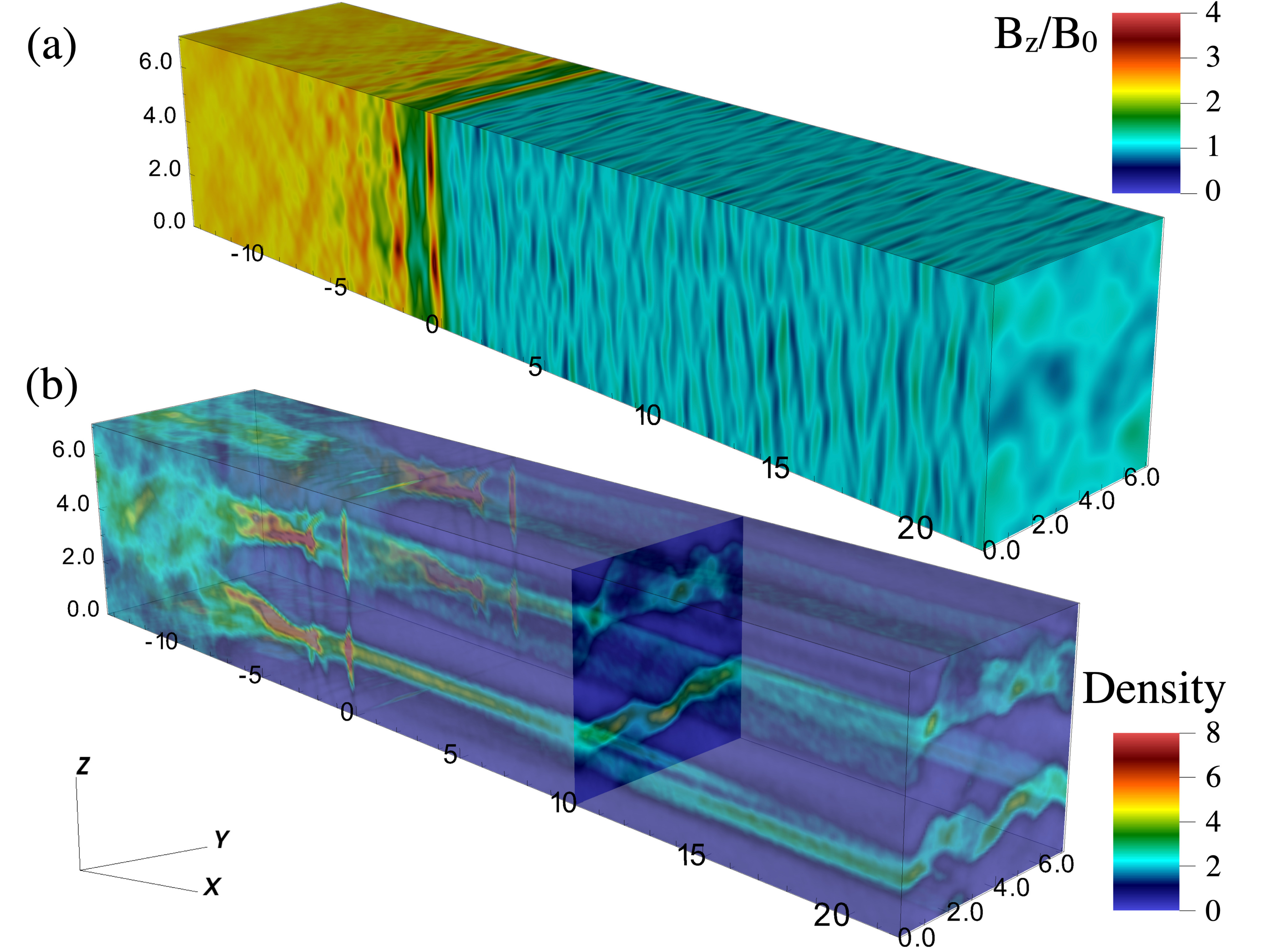}
    \caption{3D structure of magnetic field $B_z/B_0$ (top) and density (bottom, in units of $n_0$) from a simulation with $\gamma_0=3$ and $\sigma=6$ at  $\ompt=9,000$. 
    The $x$ coordinate is measured with respect to the shock location $x_{\rm sh}$, in units of $\comp$. 
    The upstream is at $x-x_{\rm sh}>0$ and the downstream at $x-x_{\rm sh}<0$. 
    In the bottom panel, the slice at $x-x_{\rm sh}=10\, \comp$ emphasizes the upstream density structures.}
    \label{fig:3d}
\end{figure}

\textit{Wave Efficiency and Spectrum.---}
\figg{3d} shows the shock structure from a simulation with $\gamma_0=3$ and $\sigma=6$, when the system has reached a quasi-steady state.
The shock exhibits a soliton-like structure with the enhanced magnetic field and density at $x\approx x_{\rm sh}$ \citep[][]{alsop_arons_88}. 
In the soliton, the incoming particles gyrate around the compressed magnetic field and form a semi-coherent ring in momentum space \citep{hoshino_91}. 
In the density cavity behind the leading soliton ($-2\,\comp\lesssim x-\xsh\lesssim 0$), the magnetic field goes back to the upstream value. 
This cavity is a peculiarity of $\sigma\gtrsim 1$ shocks \citep{plotnikov2019}. 
It controls the properties of X-mode waves, and the peak frequency of the wave spectrum corresponds to an eigenmode of the cavity. 

X-mode waves are generated by an oscillating current near the downstream side of the cavity ($x\lesssim\xsh -2\,\comp$), differently from the customary synchrotron maser description \citep{alsop_arons_88,hoshino_01}.
In \figg{3d}(a), the X-mode waves appear as ripples in $B_z$, 
 within the density cavity and in the upstream region.
Similarly, the shock emits O-mode waves appearing in $B_y$. 
Self-focusing of the precursor waves generates filamentary structures in the upstream density 
(the filamentation instability was previously studied in electron-proton unmagnetized plasma, e.g. see \citep{drake_74,sobacchi_21}).
The high magnetization inhibits particle motion across $\boldsymbol{B}_0$, so the resulting density structures appear as sheets nearly orthogonal to the pre-shock field. 
These sheets are responsible for the O-mode generation. 

\begin{figure}
\centering
    \includegraphics[width=\columnwidth]{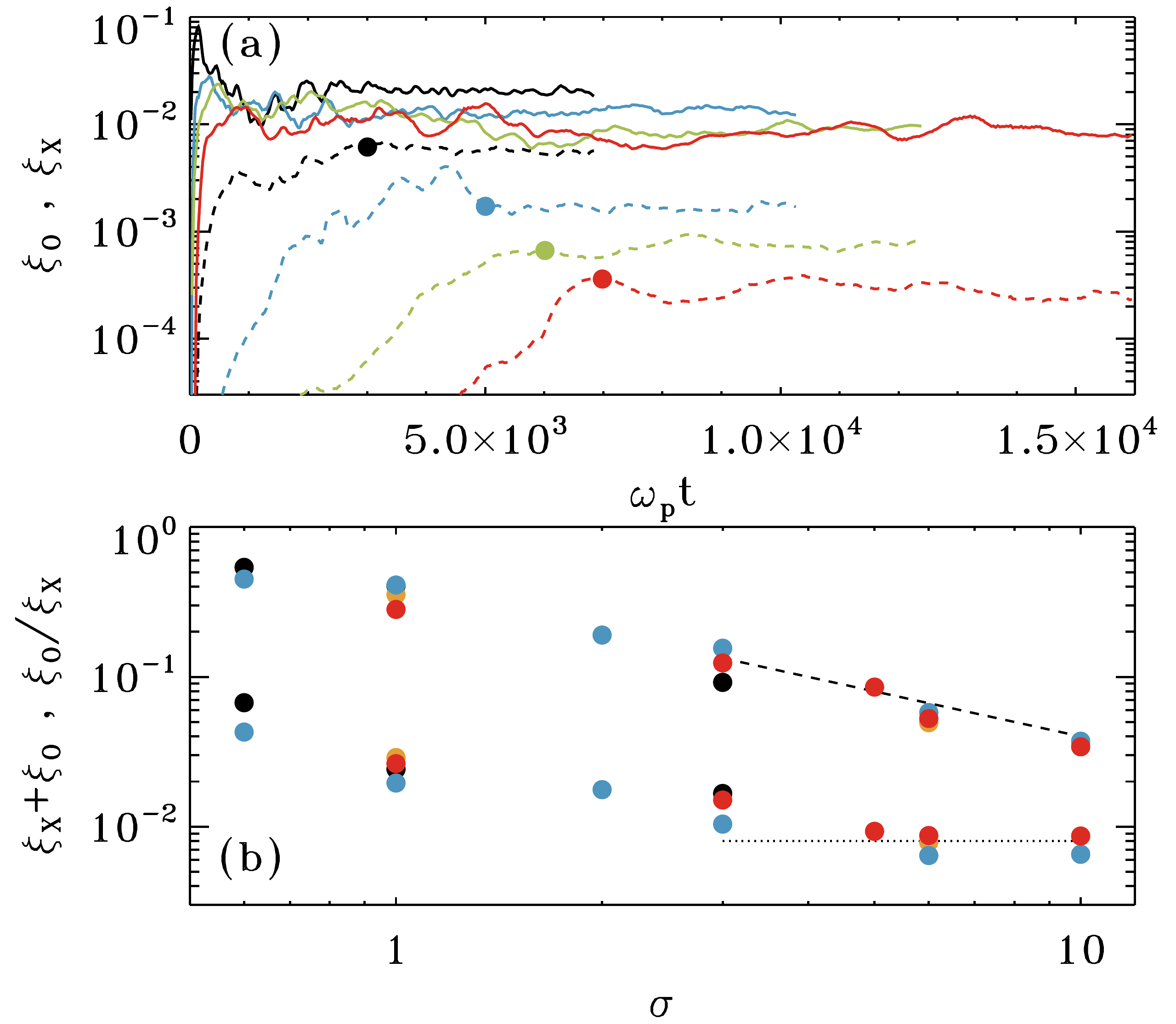}
    \caption{
    (a) Temporal evolution of the Poynting flux carried by X-modes (solid; $\xi_{\rm X}=\langle(B_z-B_0)^2\rangle/B_0^2$) and O-modes (dashed; $\xi_{\rm O}=\langle B_y^2\rangle/B_0^2$), for shocks with $\gamma_0=3$ and different magnetizations: $\sigma=1$ (black), 3 (blue), 5 (green), and 10 (red).
    The filled circles on the dashed lines mark the time when the corresponding shock approaches a quasi-steady state. 
    (b) Steady-state ratio of the O-mode and X-mode powers $\xi_{\rm O}/\xi_{\rm X}$ (upper series of points) and total Poynting flux $\xi_{\rm X}+\xi_{\rm O}$ (lower series of points), as a function of magnetization. 
    Colors indicate different choices of the upstream Lorentz factor $\gamma_0=3$ (yellow) or 10 (black), and different domain sizes $L=3.6c/\omega_{\rm p}$  (red and blue) or $7.2c/\omega_{\rm p}$ (yellow and black).
    The dashed black line indicates $\xi_{\rm O}/\xi_{\rm X}\simeq 0.4\,\sigma^{-1}$, while the dotted black line represents $\xi_{\rm X}+\xi_{\rm O}\simeq 8\times10^{-3}$.}
    \label{fig:eff}
\end{figure}

The Poynting flux carried by X-mode and O-mode waves is quantified in \figg{eff}(a), via the temporal evolution of $\xi_{\rm X}=\langle(B_z-B_0)^2\rangle/B_0^2$ and $\xi_{\rm O}=\langle B_y^2\rangle/B_0^2$ 
 for different magnetizations. 
The spatial average is taken from $5$ to 30 $\comp$ ahead of the shock. 
At late times, the Poynting fluxes settle to a steady state, and we measure their values for shocks with different $\sigma$.  
At $\sigma\gg 1$, the X-mode power asymptotes to $\xi_{\rm X}\simeq \ex{2}$, in agreement with earlier 1D and 2D results \citep{plotnikov2019,babul_sironi_20}. 
In contrast, the O-mode power drops with $\sigma$, approximately as $\xi_{\rm O}\simeq 4\times 10^{-3}\, \sigma^{-1}$. 

\begin{figure}
\centering
    \includegraphics[width=\columnwidth]{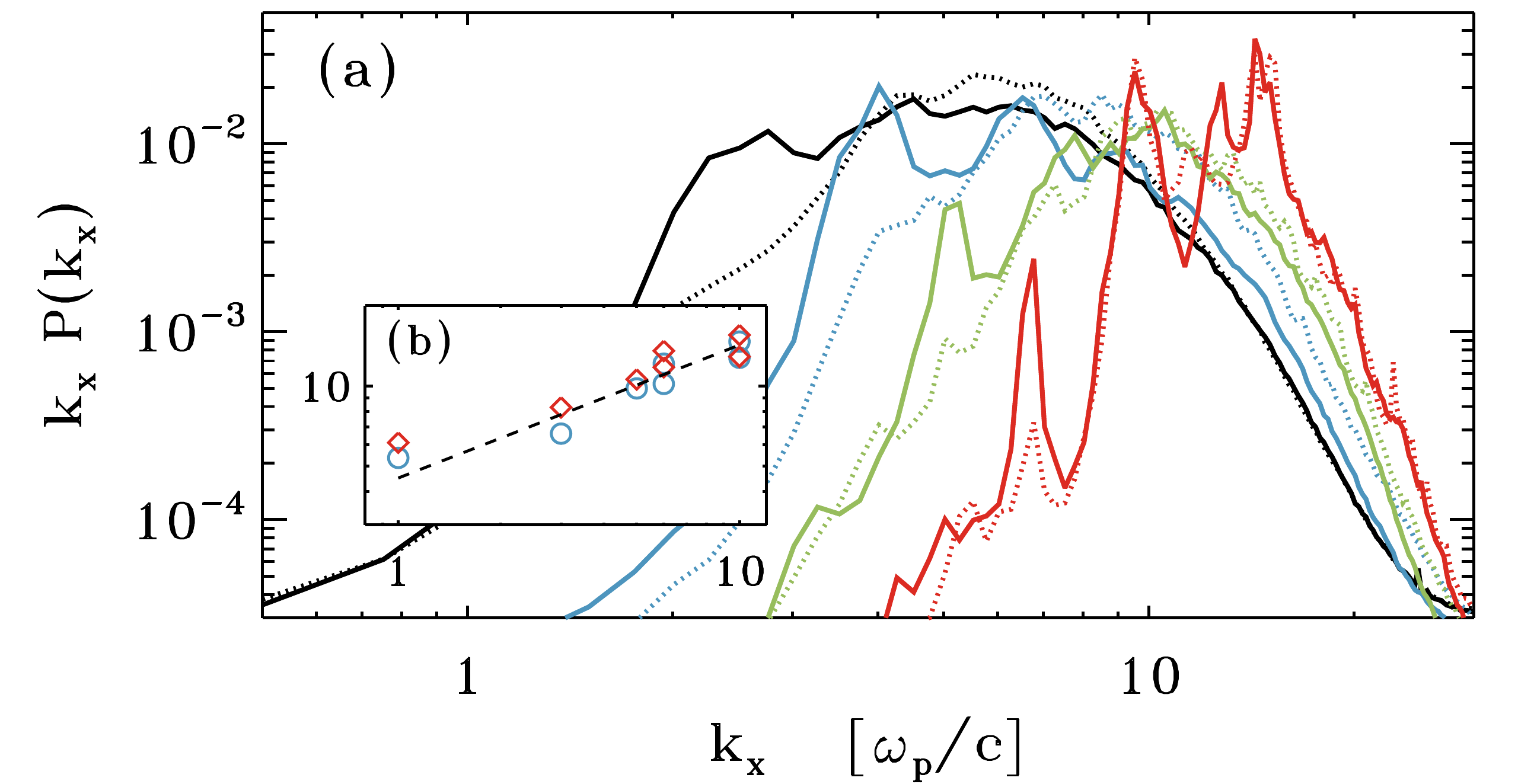}
    \caption{
1D power spectrum $k_x P(k_x)$, where $P(k_x)=\int P_{\rm 3D}(k_x,k_y,k_z)\,d k_y \,dk_z$ and $P_{\rm 3D}$ is the 3D power spectrum. 
Both X-mode  (solid) and O-mode (dotted) spectra are normalized so that $\int P(k_x)dk_x=\xi_{\rm X}$, i.e., the O-mode spectrum is shifted up by a factor $\xi_{\rm X}/\xi_{\rm O}$ to facilitate comparison with the X-mode spectrum. 
Color coding is the same as in \figg{eff}(a). 
The inset shows the energy-carrying wavenumber $\langle k_x\rangle=\int k_x P(k_x)dk_x/\int P(k_x)dk_x$, in units of $\omp/c$, as a function of magnetization (blue for X-mode, red for O-mode). 
For $\sigma=6$ and 10, \ls{the inset also shows} results from simulations with higher spatial resolution, $\comp=50$ cells, \ls{yielding similar findings}. 
The dashed black line indicates the scaling $ \langle k_x\rangle\simeq 4\sqrt{\sigma} \omp/c$.}
    \label{fig:spec}
\end{figure}

The resulting O/X mode ratio is $\xi_{\rm O}/\xi_{\rm X}\simeq 0.4\,\sigma^{-1}$ for high magnetizations (\figg{eff}(b)). 
This scaling is robust to varying the flow Lorentz factor ($\gamma_0=3$ or $10$) and domain size ($L=3.6\comp$ or $7.2c/\omega_{\rm p}$). 
The O-mode suppression with increasing $\sigma$ is consistent with previous results of 2D in-plane simulations with $\sigma \lesssim 1$ \citep{iwamoto_18}\footnote{We remark, however, that in \citet{iwamoto_18} the O-mode emission was attributed to gyro-phase bunching in Weibel-generated fields, which cannot operate in the high-$\sigma$ shocks presented in this Letter.}. 

The efficiency of the maser emission $f_{\xi}$ is
defined in the downstream frame as the fraction of incoming flow energy (electromagnetic + kinetic) converted to precursor wave energy, and we find
\be
f_{\xi}=(\xi_{\rm X}+\xi_{\rm O})\left( \frac{\sigma }{ 1+\sigma} \right) \left( \frac{1-\beta_{\rm sh} }{ \beta_0 + \beta_{\rm sh}} \right)
\label{eq:xidef}
\ee
where $\beta_{\rm sh}$ is the shock speed in units of $c$. 
While the ratio
$\xi_{\rm O}/\xi_{\rm X}$ monotonically drops with $\sigma$, the overall wave energy flux asymptotes to a constant value $\xi_{\rm X}+\xi_{\rm O}\sim \xi_{\rm X}\simeq 0.01$ for $\sigma\gg1$, and gives
$f_\xi\simeq \ex{3}\,\sigma^{-1}$. 
In the shock maser scenario for FRBs, this quantifies the fraction of the blast wave energy that is converted into FRB energy. 
The quantity $\xi_{\rm X}+\xi_{\rm O}$ also determines the dimensionless strength parameter of precursor waves at $\sigma\gg 1$:  
$a\sim 0.3 \gamma_0\sqrt{\xi_{\rm X}+\xi_{\rm O}}\simeq 0.03\,\gamma_0$ \citep{plotnikov2019}. 
For ultra-relativistic shocks the strength parameter can  exceed unity.

The spectrum of precursor waves is presented in \figg{spec} for different magnetizations. 
The spectra are computed in the post-shock frame, 
from the same upstream region ($5\,\comp< x-\xsh<30\,\comp$) where we extracted the precursor efficiency. 
Both X-mode and O-mode spectra peak at higher wavenumbers for larger magnetizations. 
This is also illustrated by the dependence on $\sigma$ of the energy-carrying wavenumber $\langle k_x\rangle\simeq 4\sqrt{\sigma} \omp/c$ (inset in \figg{spec}(b)). 
\figg{spec} demonstrates that at high wavenumbers ($k_x/\langle k_x\rangle\gtrsim 1$) X-mode and O-mode spectra are similar. 
They differ, however, at lower wavenumbers; 
the spectral feature at $k_x\simeq 0.5 \langle k_x\rangle$ only appears in the X-mode. 
Near this wavenumber, the O/X mode ratio is one order of magnitude below the average $\xi_{\rm O}/\xi_{\rm X}$.

\begin{figure}
\centering
    \includegraphics[width=\columnwidth]{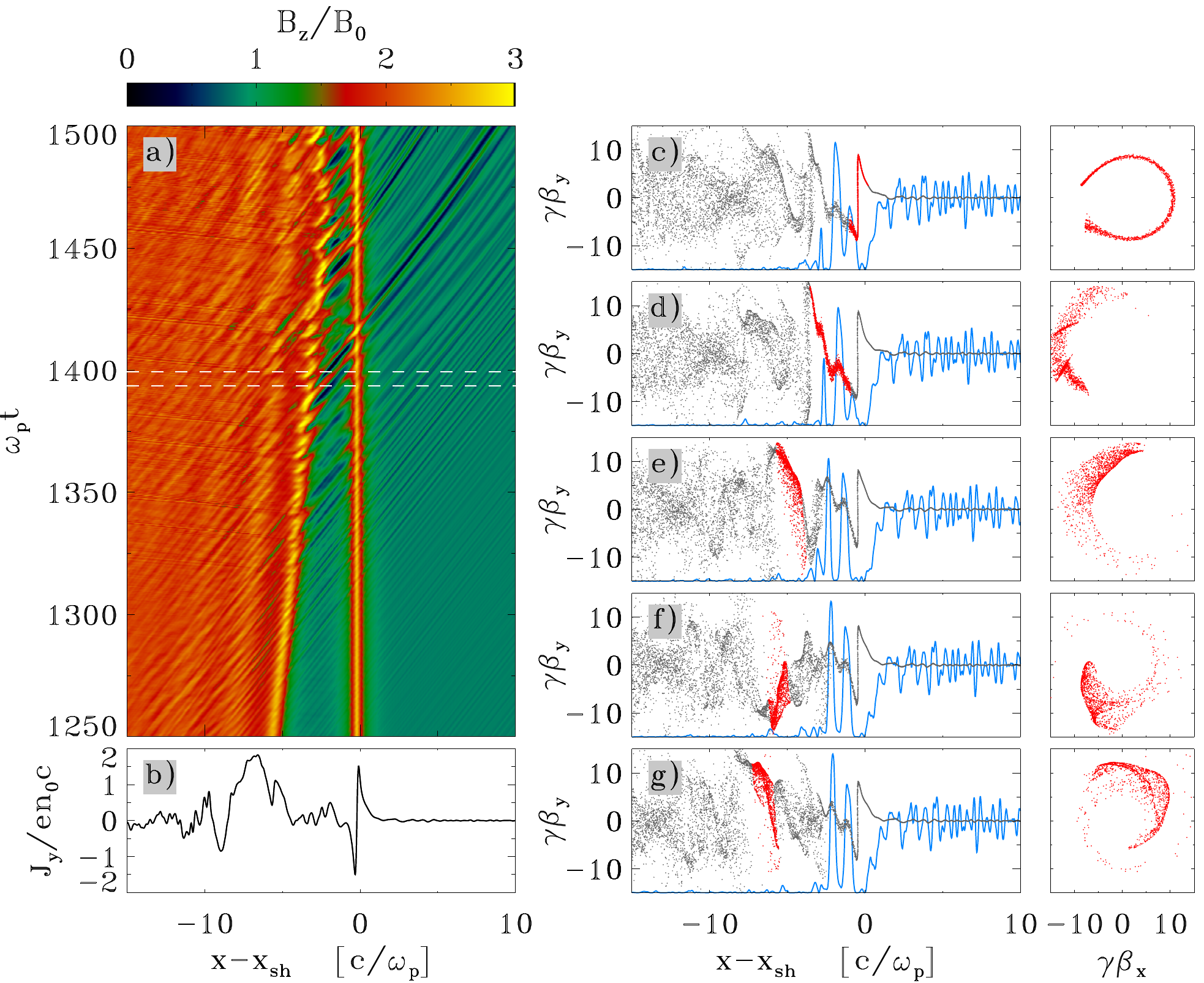}
    \caption{ 
    Shock ``breathing,'' as observed in the simpler 1D simulation ($\gamma_0=10$, $\sigma=3$), which fully captures the X-mode generation.
    (a) Spatio-temporal structure of $B_z/B_0$. (b)  Electric current density $J_y/en_0c$ at $\ompt=1400$ (the moment indicated by the upper dashed line in panel (a)). 
    (c)-(g) Sequence of five snapshots for randomly sampled positrons in the $x-\gamma\beta_y$ phase space  and  for a selected subset of positrons (in red) in the $\gamma\beta_x-\gamma\beta_y$ momentum space. 
The snapshots cover the time interval 
$1393<\ompt<1400$
(indicated by the two horizontal dashed lines in (a)).
Positrons highlighted in red are selected so that at $\ompt=1400$ they contribute to the current peak at $-10\,\comp\lesssim x-\xsh \lesssim-5\,\comp$ in  (b). The blue lines show $(E^2-B^2)/(E^2+B^2)$ (for these curves, the vertical boundaries correspond to -1 and 1).
}
    \label{fig:xmode}
\end{figure}

\textit{X-Mode Emission Mechanism.---}
The X-mode generation 
is well captured in 1D models. One example
with $\gamma_0=10$ and $\sigma=3$
is shown
in \figg{xmode}. 
The X-mode waves appear as ripples in $B_z$ propagating into the upstream region, after the width of the shock cavity has settled into a steady state ($\ompt\gtrsim 1350$). 
The current near the leading soliton then remains nearly time-independent. The oscillating current $J_y$ emitting the X-mode waves 
is localized downstream of the {\em second} soliton, at $-10\,\comp\lesssim x-\xsh \lesssim-5\,\comp$ (\figg{xmode}(b)).
The generated waves
propagate both toward the downstream (where they are eventually absorbed) and through the density cavity into the upstream.

In the shock frame, which moves with Lorentz factor $\approx\sqrt{\sigma}$ relative to the downstream, the field satisfies the jump condition $B_{d,\rm sh}/B_{u,\rm sh}=1+1/2\sigma$ \citep[][]{petri_lyubarsky_07}.
Here, $B_{u,\rm sh}\simeq 2\sqrt{\sigma}B_0$ and $B_{d,\rm sh}$ are the upstream and downstream fields, respectively. 
The 
net
surface current of the multiple soliton structure of the shock is $\Sigma\sim c(B_{d,\rm sh}-B_{u,\rm sh})/4\pi\sim c B_0/(4 \pi \sqrt{\sigma})$. 
Assume that the surface current varies with some amplitude $\zeta=\delta \Sigma / \Sigma$.
In the shock frame, the emitted wave has the amplitude $\delta B_{\rm sh}\sim \zeta B_0/(2\sqrt{\sigma})$. 
This wave, viewed in the downstream frame of the simulation, has $\delta B\approx \sqrt{\sigma}\, \delta B_{\rm sh}$, which gives the precursor with amplitude $\sim \zeta B_0$ and $\xi_{\rm X}\approx \zeta^2$. 
Thus, the constant $\xi_{\rm X}\approx 10^{-2}$ observed at $\sigma\gg 1$ is consistent with the constant amplitude $\zeta\approx 0.1$ of the shock fluctuations. 

Most positrons in the 
wave emission region ($-10\,\comp<x-x_{\rm sh}<-5\,\comp$) have $\gby>0$. 
By symmetry of the 1D simulation, at any given location electrons and positrons have the same densities and $\gamma\beta_x$, and opposite $\gamma\beta_y$, which determines the
 current
$J_y$.
\figg{xmode} shows the history of the positrons ending up at $-10\,\comp\lesssim x-\xsh \lesssim-5\,\comp$.
At the shock, they initially form a ring in momentum space
\citep[][]{gallant_92}. 
Their subsequent 
motion through the density cavity 
is affected by the waves 
generated by earlier generations of shocked particles.
The waves create (by
linear superposition) 
regions of 
$E>B$ (see the blue lines in \figg{xmode}). 
The $E>B$ region extends across roughly half a wavelength. 

Particles exposed to $E>B$ get accelerated, at a rate that depends on the wave amplitude and the particle $\beta_y$.
This wave-particle interaction increases the energy spread of incoming particles, so their arc in the $\gbx-\gby$ 
space grows in radius and thickness, as they move past the leading soliton (compare \figg{xmode}(d) and (c)). 
The particles at the inner edge of the arc 
(lower $\gamma$) 
gyrate faster than the particles at the outer edge (higher $\gamma$). 
Initially, lower-energy positrons lag in phase behind higher-energy ones (\figg{xmode}(e); positrons move clockwise in the $\gamma\beta_x-\gamma\beta_y$ plane),
but later catch up
due to their 
shorter gyroperiod. 
This creates gyro-phase bunching (\figg{xmode}(f) and (g)),
and produces 
an intermittent enhancement of
current density $J_y$. 
\ls{The current
oscillates on the gyroperiod of post-shock particles and its harmonics, generating the X-mode}.

Without wave activity, the magnetic field in the cavity is $\boldsymbol{B}_0$. 
The condition $E>B$ is realized where the wave has $\delta B_z<0$ and $\delta E_y<0$, so $|B_z|=|B_0+\delta B_z|<B_0$ and $|E_y|=|-E_0+\delta E_y|>E_0$. 
Since $E_y<0$, positrons are accelerated in $E>B$ regions if their $\beta_y<0$ (conversely, $\beta_y>0$ for electrons). 
This condition occurs in half of the cavity, just behind the leading soliton. 
Therefore, the process of wave generation is self-reinforced if the half-wavelength of the X-mode 
is approximately equal to the half-thickness of the cavity.
This explains why the peak frequency of the wave spectrum corresponds to an eigenmode of the cavity \citep{plotnikov2019}.

Thus, X-mode waves are generated by a non-local positive feedback loop: 
(\textit{i}) in the density cavity, waves propagating upstream lead to $E>B$, which perturbs the energies of fresh particles entering the cavity from the shock front; 
(\textit{ii}) higher-energy particles gyrate slower than lower-energy ones, leading to gyro-phase bunching;  
(\textit{iii}) this produces a net current oscillating on the particle gyration time, leading to more wave production. \ls{The non-locality of the feedback loop differentiates our mechanism from the standard (local) description of the synchrotron maser. Also, the precursor emission in high-$\sigma$ shocks cannot be attributed to the standard maser mechanism, since a seed wave cannot be considerably amplified while particles cross the shock \citep{lyubarsky_21}.}

\begin{figure}
\centering
    \includegraphics[width=\columnwidth]{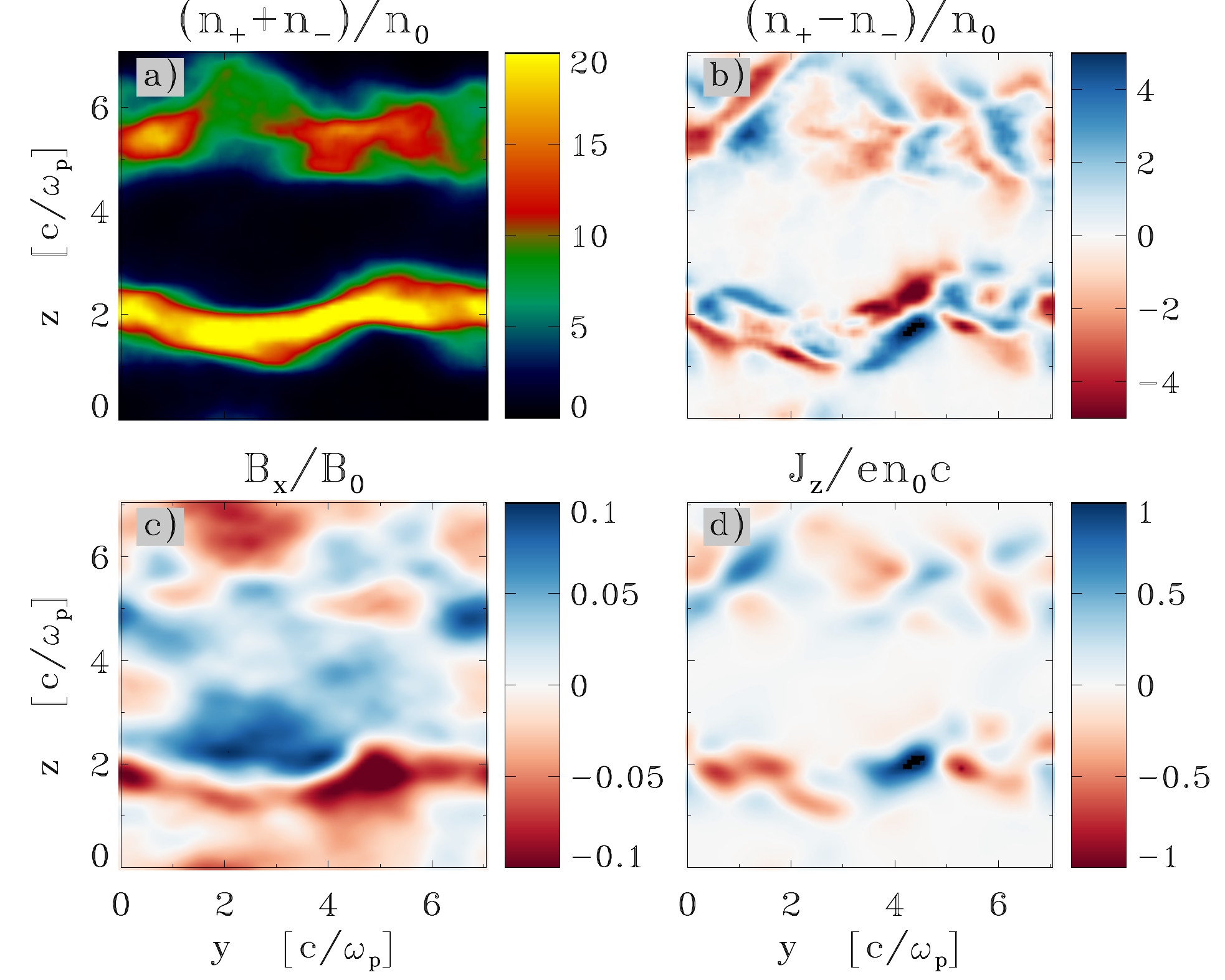}
    \caption{{Charge separation and field line bending (from the 
      3D simulation presented in \figg{3d}), explaining the O-mode generation.
    At the location 
    of the leading soliton $x=\xsh$, we show 2D $y-z$ slices of: (a) $n_{+}+n_{-}$, where $n_{+}$ and $n_{-}$ are the positron and electron number densities; (b)  $n_{+}-n_{-}$, as a proxy for 
     charge separation; (c)  $B_x/B_0$; (d) 
     current density $J_z/en_0c$.}}
    \label{fig:omode}
\end{figure}

\textit{O-Mode Emission Mechanism.---}
The physics of the O-mode generation is inherently 3D.   
In \figg{omode}, we employ the same 3D simulation of \figg{3d}, and show 2D slices at the location of the leading soliton, $x=\xsh$ 
\footnote{Even though we focus on the leading soliton, the features described below are observed throughout the density cavity until the second soliton.}.

The O-mode-generating current $J_z$ is ultimately related to the sheet-like density layers  produced by self-focusing of the precursor waves (\figg{omode}(a)). 
Significant charge separation develops at the boundaries of the density sheets (\figg{omode}(b)), because positrons and electrons flowing into the shock gyrate in opposite directions. 
This leads to charge separation as long as the sheets are not invariant along the $y$ direction perpendicular to the initial field, e.g., for tilted or inhomogeneous sheets.

If only the $B_z$ component was present, the charge bunches would move in the $x-y$ plane, and would not generate any $J_z$. 
A nonzero $B_x$ appears at the shock because the non-uniform ram pressure of the sheets causes field-line bending in the $x$-$z$ plane. 
Its energy density 
scales with the incoming kinetic energy density, $B_x^2/B_0^2\propto \gamma_0 n_0 m c^2/B_0^2\propto\sigma^{-1}$.
Thus, the field lines near the front are no longer perpendicular to the flow velocity, and the charge bunches slide along the field, developing a small 
$\beta_z\simeq\beta_\parallel\simeq \sqrt{B_x^2/B_0^2}$, which in turn leads to the O-mode-generating $J_z$  (\figg{omode}(d)).
This 
implies that the O/X mode power ratio should scale as $\xi_{\rm O}/\xi_{\rm X}\propto \beta_z^2\propto B_x^2/B_0^2\propto \sigma^{-1}$, as observed in \figg{eff}(b).

In summary, O-mode generation can be properly captured only in 3D. It requires breaking the symmetry both (\textit{i}) along $y$ --- 
enabling 
charge separation at the boundaries of the density sheets when the incoming particles begin to gyrate in the shock-compressed field; and (\textit{ii}) along 
$z$ 
($\boldsymbol{B}_0$ 
direction) --- enabling $B_x$ generation via field-line bending
by
the high-density sheets colliding with the shock. The charge bunches slide along the perturbed field lines, creating the variable O-mode current $J_z$.

\textit{Summary.---}
By means of 3D PIC simulations, we have characterized O-mode and X-mode waves emitted by relativistic magnetized shocks propagating in magnetically-dominated ($\sigma>1$) pair plasmas. 
{The fraction of incoming energy converted into precursor waves is $f_\xi\simeq \ex{3} \, \sigma^{-1}$, and the energy-carrying wavenumber is  $\langle k_x \rangle\simeq 4 \,\omega_{\rm c}/c$. 
The O/X mode power ratio is $\xi_{\rm O}/\xi_{\rm X}\simeq 0.4\,\sigma^{-1}$, regardless of the shock Lorentz factor. 
While O-mode and X-mode spectra overlap at high wavenumbers, the narrow spectral feature at $k_x\simeq 0.5 \langle k_x \rangle$ is much stronger in the X-mode.}


\ls{
Our results provide important plasma-physical inputs for FRB emission models, by demonstrating that high-$\sigma$ shocks can emit electromagnetic waves with a high degree of linear polarization, as observed in some FRBs \citep[][]{michilli_18}.
}
By calculating the power-weighted Stokes' parameters for a line of sight along the shock normal, one can compute the degree of linear polarization $P=Q/I$ \citep[][$U=V=0$ is well satisfied in our simulations]{rybicki_lightman_79} intrinsic to the shock emission \footnote{
Detailed calculations
of the observed polarization degree from FRB-producing shocks, accounting for both shock curvature and propagation effects, will be presented elsewhere.}.
Since $Q=\xi_{\rm X}-\xi_{\rm O}$ and $I=\xi_{\rm X}+\xi_{\rm O}$, the degree of linear polarization for $\xi_{\rm O}/\xi_{\rm X}\ll1$  is $P\simeq 1-2\xi_{\rm O}/\xi_{\rm X}\simeq 1-0.8\sigma^{-1}$.

\phantom{xx}
\begin{acknowledgments}
  We thank E. Sobacchi and N. Sridhar for useful comments. L.S. acknowledges support from the Sloan Fellowship, the Cottrell Scholars Award, DoE DE-SC0016542, NASA 80NSSC18K1104 and NSF PHY-1903412. A.M.B. acknowledges support by NASA grant NNX 17AK37G, NSF grant AST 2009453, Simons Foundation grant \#446228, and the Humboldt Foundation.  The simulations have been performed at Columbia (Habanero and Terremoto), with NERSC (Cori) and NASA (Pleiades) resources.
\end{acknowledgments}

\section{Supplementary Material: Numerical Simulation Parameters}
The plasma skin depth in our 3D simulations is well resolved with $\comp=25$ cells, to capture the high-frequency wave spectrum and mitigate the numerical Cerenkov instability \citep{iwamoto_17}. 
For $\sigma=6$ and $10$, we also present results with doubled resolution, $\comp=50$ cells. \ls{As shown in the main text, the energy-carrying wavenumber of the wave spectrum is  $\simeq 4 \,\sqrt{\sigma}\,\omega_{\rm p}/c$, so a careful assessment of convergence with respect to spatial resolution is particularly important at high magnetizations.}
We vary the transverse box size $L$ between 3.6 and $7.2\,\comp$, which we find sufficient to capture multi-dimensional effects. 
We employ periodic boundary conditions in $y$ and $z$. 
The upstream plasma has 3 particles per cell per species, but we have tested that even 1 particle per cell gives converged results.
The numerical speed of light is $0.45\,{\rm cells/timestep}$. 
Our longest simulations extend up to $\sim 16,000\,\omp^{-1}$ ($\sim900,000$ timesteps).

The simulations are optimized in $x$ to enable longer numerical experiments.
The incoming plasma is injected by a ``moving injector,'' which moves along $+{\hat{x}}$ at the speed of light, thus retaining all regions in causal contact with the shock. 
The box is expanded in $x$ as the injector approaches the right boundary \citep[][]{sironi_spitkovsky_09}. 
The left edge of the downstream region (hereafter, the ``wall'') acts as a particle reflector and provides conducting boundary conditions for the electromagnetic fields. 
It is initially located at $x=0$, but gets periodically relocated closer to the shock to save memory and computational time;
we ensure a minimum distance of $\sim 100\,\comp$ between the wall and the shock.
We also performed simulations with a stationary wall and verified that the wall relocation does not influence the shock or the precursor waves \citep[][]{babul_sironi_20}.

We also present one 1D simulation with $\gamma_0=10$ and $\sigma=3$ to clarify the physics of X-mode emission. 
The 1D run resolves the skin depth with 50 cells and is initialized with 40 particles per cell.



\begin{thebibliography}{32}
\expandafter\ifx\csname natexlab\endcsname\relax\def\natexlab#1{#1}\fi
\expandafter\ifx\csname bibnamefont\endcsname\relax
  \def\bibnamefont#1{#1}\fi
\expandafter\ifx\csname bibfnamefont\endcsname\relax
  \def\bibfnamefont#1{#1}\fi
\expandafter\ifx\csname citenamefont\endcsname\relax
  \def\citenamefont#1{#1}\fi
\expandafter\ifx\csname url\endcsname\relax
  \def\url#1{\texttt{#1}}\fi
\expandafter\ifx\csname urlprefix\endcsname\relax\def\urlprefix{URL }\fi
\providecommand{\bibinfo}[2]{#2}
\providecommand{\eprint}[2][]{\url{#2}}

\bibitem[{\citenamefont{{Petroff} et~al.}(2019)\citenamefont{{Petroff},
  {Hessels}, and {Lorimer}}}]{petroff_19}
\bibinfo{author}{\bibfnamefont{E.}~\bibnamefont{{Petroff}}},
  \bibinfo{author}{\bibfnamefont{J.~W.~T.} \bibnamefont{{Hessels}}},
  \bibnamefont{and} \bibinfo{author}{\bibfnamefont{D.~R.}
  \bibnamefont{{Lorimer}}}, \bibinfo{journal}{\aapr}
  \textbf{\bibinfo{volume}{27}}, \bibinfo{eid}{4} (\bibinfo{year}{2019}),
  \eprint{1904.07947}.

\bibitem[{\citenamefont{{Cordes} and {Chatterjee}}(2019)}]{cordes_19}
\bibinfo{author}{\bibfnamefont{J.~M.} \bibnamefont{{Cordes}}} \bibnamefont{and}
  \bibinfo{author}{\bibfnamefont{S.}~\bibnamefont{{Chatterjee}}},
  \bibinfo{journal}{\araa} \textbf{\bibinfo{volume}{57}}, \bibinfo{pages}{417}
  (\bibinfo{year}{2019}), \eprint{1906.05878}.

\bibitem[{\citenamefont{{Platts} et~al.}(2019)\citenamefont{{Platts},
  {Weltman}, {Walters}, {Tendulkar}, {Gordin}, and {Kandhai}}}]{platts_19}
\bibinfo{author}{\bibfnamefont{E.}~\bibnamefont{{Platts}}},
  \bibinfo{author}{\bibfnamefont{A.}~\bibnamefont{{Weltman}}},
  \bibinfo{author}{\bibfnamefont{A.}~\bibnamefont{{Walters}}},
  \bibinfo{author}{\bibfnamefont{S.~P.} \bibnamefont{{Tendulkar}}},
  \bibinfo{author}{\bibfnamefont{J.~E.~B.} \bibnamefont{{Gordin}}},
  \bibnamefont{and}
  \bibinfo{author}{\bibfnamefont{S.}~\bibnamefont{{Kandhai}}},
  \bibinfo{journal}{\physrep} \textbf{\bibinfo{volume}{821}},
  \bibinfo{pages}{1} (\bibinfo{year}{2019}), \eprint{1810.05836}.

\bibitem[{\citenamefont{{Lyubarsky}}(2021)}]{lyubarsky_21}
\bibinfo{author}{\bibfnamefont{Y.}~\bibnamefont{{Lyubarsky}}},
  \bibinfo{journal}{Universe} \textbf{\bibinfo{volume}{7}}, \bibinfo{pages}{56}
  (\bibinfo{year}{2021}), \eprint{2103.00470}.

\bibitem[{\citenamefont{{Katz}}(2016)}]{katz_16}
\bibinfo{author}{\bibfnamefont{J.~I.} \bibnamefont{{Katz}}},
  \bibinfo{journal}{Modern Physics Letters A} \textbf{\bibinfo{volume}{31}},
  \bibinfo{eid}{1630013} (\bibinfo{year}{2016}), \eprint{1604.01799}.

\bibitem[{\citenamefont{{Scholz} and {Chime/Frb
  Collaboration}}(2020)}]{scholz_20}
\bibinfo{author}{\bibfnamefont{P.}~\bibnamefont{{Scholz}}} \bibnamefont{and}
  \bibinfo{author}{\bibnamefont{{Chime/Frb Collaboration}}},
  \bibinfo{journal}{The Astronomer's Telegram}
  \textbf{\bibinfo{volume}{13681}}, \bibinfo{pages}{1} (\bibinfo{year}{2020}).

\bibitem[{\citenamefont{{Bochenek} et~al.}(2020)\citenamefont{{Bochenek},
  {Ravi}, {Belov}, {Hallinan}, {Kocz}, {Kulkarni}, and
  {McKenna}}}]{bochenek_20}
\bibinfo{author}{\bibfnamefont{C.~D.} \bibnamefont{{Bochenek}}},
  \bibinfo{author}{\bibfnamefont{V.}~\bibnamefont{{Ravi}}},
  \bibinfo{author}{\bibfnamefont{K.~V.} \bibnamefont{{Belov}}},
  \bibinfo{author}{\bibfnamefont{G.}~\bibnamefont{{Hallinan}}},
  \bibinfo{author}{\bibfnamefont{J.}~\bibnamefont{{Kocz}}},
  \bibinfo{author}{\bibfnamefont{S.~R.} \bibnamefont{{Kulkarni}}},
  \bibnamefont{and} \bibinfo{author}{\bibfnamefont{D.~L.}
  \bibnamefont{{McKenna}}}, \bibinfo{journal}{\nat}
  \textbf{\bibinfo{volume}{587}}, \bibinfo{pages}{59} (\bibinfo{year}{2020}),
  \eprint{2005.10828}.

\bibitem[{\citenamefont{{Beloborodov}}(2017)}]{belo_17}
\bibinfo{author}{\bibfnamefont{A.~M.} \bibnamefont{{Beloborodov}}},
  \bibinfo{journal}{\apjl} \textbf{\bibinfo{volume}{843}}, \bibinfo{eid}{L26}
  (\bibinfo{year}{2017}), \eprint{1702.08644}.

\bibitem[{\citenamefont{{Metzger} et~al.}(2019)\citenamefont{{Metzger},
  {Margalit}, and {Sironi}}}]{lorenzometzger}
\bibinfo{author}{\bibfnamefont{B.~D.} \bibnamefont{{Metzger}}},
  \bibinfo{author}{\bibfnamefont{B.}~\bibnamefont{{Margalit}}},
  \bibnamefont{and} \bibinfo{author}{\bibfnamefont{L.}~\bibnamefont{{Sironi}}},
  \bibinfo{journal}{\mnras} \textbf{\bibinfo{volume}{485}},
  \bibinfo{pages}{4091} (\bibinfo{year}{2019}), \eprint{1902.01866}.

\bibitem[{\citenamefont{{Beloborodov}}(2020)}]{belo_20}
\bibinfo{author}{\bibfnamefont{A.~M.} \bibnamefont{{Beloborodov}}},
  \bibinfo{journal}{\apj} \textbf{\bibinfo{volume}{896}}, \bibinfo{eid}{142}
  (\bibinfo{year}{2020}), \eprint{1908.07743}.

\bibitem[{\citenamefont{{Yuan} et~al.}(2020)\citenamefont{{Yuan},
  {Beloborodov}, {Chen}, and {Levin}}}]{yuan_20}
\bibinfo{author}{\bibfnamefont{Y.}~\bibnamefont{{Yuan}}},
  \bibinfo{author}{\bibfnamefont{A.~M.} \bibnamefont{{Beloborodov}}},
  \bibinfo{author}{\bibfnamefont{A.~Y.} \bibnamefont{{Chen}}},
  \bibnamefont{and} \bibinfo{author}{\bibfnamefont{Y.}~\bibnamefont{{Levin}}},
  \bibinfo{journal}{\apjl} \textbf{\bibinfo{volume}{900}}, \bibinfo{eid}{L21}
  (\bibinfo{year}{2020}), \eprint{2006.04649}.

\bibitem[{\citenamefont{{Alsop} and {Arons}}(1988)}]{alsop_arons_88}
\bibinfo{author}{\bibfnamefont{D.}~\bibnamefont{{Alsop}}} \bibnamefont{and}
  \bibinfo{author}{\bibfnamefont{J.}~\bibnamefont{{Arons}}},
  \bibinfo{journal}{Physics of Fluids} \textbf{\bibinfo{volume}{31}},
  \bibinfo{pages}{839} (\bibinfo{year}{1988}).

\bibitem[{\citenamefont{{Hoshino} and {Arons}}(1991)}]{hoshino_91}
\bibinfo{author}{\bibfnamefont{M.}~\bibnamefont{{Hoshino}}} \bibnamefont{and}
  \bibinfo{author}{\bibfnamefont{J.}~\bibnamefont{{Arons}}},
  \bibinfo{journal}{Physics of Fluids B} \textbf{\bibinfo{volume}{3}},
  \bibinfo{pages}{818} (\bibinfo{year}{1991}).

\bibitem[{Note1()}]{Note1}
Note1, \bibinfo{note}{the generation of coherent emission is not directly due
  to wave amplification via a maser process, but still we shall refer to this
  as the ``synchrotron maser,'' because this term is widely used in the
  literature.}

\bibitem[{\citenamefont{{Michilli} et~al.}(2018)\citenamefont{{Michilli},
  {Seymour}, {Hessels}, {Spitler}, {Gajjar}, {Archibald}, {Bower},
  {Chatterjee}, {Cordes}, {Gourdji} et~al.}}]{michilli_18}
\bibinfo{author}{\bibfnamefont{D.}~\bibnamefont{{Michilli}}},
  \bibinfo{author}{\bibfnamefont{A.}~\bibnamefont{{Seymour}}},
  \bibinfo{author}{\bibfnamefont{J.~W.~T.} \bibnamefont{{Hessels}}},
  \bibinfo{author}{\bibfnamefont{L.~G.} \bibnamefont{{Spitler}}},
  \bibinfo{author}{\bibfnamefont{V.}~\bibnamefont{{Gajjar}}},
  \bibinfo{author}{\bibfnamefont{A.~M.} \bibnamefont{{Archibald}}},
  \bibinfo{author}{\bibfnamefont{G.~C.} \bibnamefont{{Bower}}},
  \bibinfo{author}{\bibfnamefont{S.}~\bibnamefont{{Chatterjee}}},
  \bibinfo{author}{\bibfnamefont{J.~M.} \bibnamefont{{Cordes}}},
  \bibinfo{author}{\bibfnamefont{K.}~\bibnamefont{{Gourdji}}},
  \bibnamefont{et~al.}, \bibinfo{journal}{\nat} \textbf{\bibinfo{volume}{553}},
  \bibinfo{pages}{182} (\bibinfo{year}{2018}), \eprint{1801.03965}.

\bibitem[{\citenamefont{{Langdon} et~al.}(1988)\citenamefont{{Langdon},
  {Arons}, and {Max}}}]{langdon_88}
\bibinfo{author}{\bibfnamefont{A.~B.} \bibnamefont{{Langdon}}},
  \bibinfo{author}{\bibfnamefont{J.}~\bibnamefont{{Arons}}}, \bibnamefont{and}
  \bibinfo{author}{\bibfnamefont{C.~E.} \bibnamefont{{Max}}},
  \bibinfo{journal}{Physical Review Letters} \textbf{\bibinfo{volume}{61}},
  \bibinfo{pages}{779} (\bibinfo{year}{1988}).

\bibitem[{\citenamefont{{Gallant} et~al.}(1992)\citenamefont{{Gallant},
  {Hoshino}, {Langdon}, {Arons}, and {Max}}}]{gallant_92}
\bibinfo{author}{\bibfnamefont{Y.~A.} \bibnamefont{{Gallant}}},
  \bibinfo{author}{\bibfnamefont{M.}~\bibnamefont{{Hoshino}}},
  \bibinfo{author}{\bibfnamefont{A.~B.} \bibnamefont{{Langdon}}},
  \bibinfo{author}{\bibfnamefont{J.}~\bibnamefont{{Arons}}}, \bibnamefont{and}
  \bibinfo{author}{\bibfnamefont{C.~E.} \bibnamefont{{Max}}},
  \bibinfo{journal}{\apj} \textbf{\bibinfo{volume}{391}}, \bibinfo{pages}{73}
  (\bibinfo{year}{1992}).

\bibitem[{\citenamefont{{Sironi} and
  {Spitkovsky}}(2009)}]{sironi_spitkovsky_09}
\bibinfo{author}{\bibfnamefont{L.}~\bibnamefont{{Sironi}}} \bibnamefont{and}
  \bibinfo{author}{\bibfnamefont{A.}~\bibnamefont{{Spitkovsky}}},
  \bibinfo{journal}{\apj} \textbf{\bibinfo{volume}{698}}, \bibinfo{pages}{1523}
  (\bibinfo{year}{2009}), \eprint{0901.2578}.

\bibitem[{\citenamefont{{Iwamoto} et~al.}(2017)\citenamefont{{Iwamoto},
  {Amano}, {Hoshino}, and {Matsumoto}}}]{iwamoto_17}
\bibinfo{author}{\bibfnamefont{M.}~\bibnamefont{{Iwamoto}}},
  \bibinfo{author}{\bibfnamefont{T.}~\bibnamefont{{Amano}}},
  \bibinfo{author}{\bibfnamefont{M.}~\bibnamefont{{Hoshino}}},
  \bibnamefont{and}
  \bibinfo{author}{\bibfnamefont{Y.}~\bibnamefont{{Matsumoto}}},
  \bibinfo{journal}{\apj} \textbf{\bibinfo{volume}{840}}, \bibinfo{eid}{52}
  (\bibinfo{year}{2017}), \eprint{1704.04411}.

\bibitem[{\citenamefont{{Iwamoto} et~al.}(2018)\citenamefont{{Iwamoto},
  {Amano}, {Hoshino}, and {Matsumoto}}}]{iwamoto_18}
\bibinfo{author}{\bibfnamefont{M.}~\bibnamefont{{Iwamoto}}},
  \bibinfo{author}{\bibfnamefont{T.}~\bibnamefont{{Amano}}},
  \bibinfo{author}{\bibfnamefont{M.}~\bibnamefont{{Hoshino}}},
  \bibnamefont{and}
  \bibinfo{author}{\bibfnamefont{Y.}~\bibnamefont{{Matsumoto}}},
  \bibinfo{journal}{\apj} \textbf{\bibinfo{volume}{858}}, \bibinfo{eid}{93}
  (\bibinfo{year}{2018}), \eprint{1803.10027}.

\bibitem[{\citenamefont{{Plotnikov} et~al.}(2018)\citenamefont{{Plotnikov},
  {Grassi}, and {Grech}}}]{plotnikov_18}
\bibinfo{author}{\bibfnamefont{I.}~\bibnamefont{{Plotnikov}}},
  \bibinfo{author}{\bibfnamefont{A.}~\bibnamefont{{Grassi}}}, \bibnamefont{and}
  \bibinfo{author}{\bibfnamefont{M.}~\bibnamefont{{Grech}}},
  \bibinfo{journal}{\mnras} \textbf{\bibinfo{volume}{477}},
  \bibinfo{pages}{5238} (\bibinfo{year}{2018}), \eprint{1712.02883}.

\bibitem[{\citenamefont{{Plotnikov} and {Sironi}}(2019)}]{plotnikov2019}
\bibinfo{author}{\bibfnamefont{I.}~\bibnamefont{{Plotnikov}}} \bibnamefont{and}
  \bibinfo{author}{\bibfnamefont{L.}~\bibnamefont{{Sironi}}},
  \bibinfo{journal}{\mnras} \textbf{\bibinfo{volume}{485}},
  \bibinfo{pages}{3816} (\bibinfo{year}{2019}), \eprint{1901.01029}.

\bibitem[{\citenamefont{{Babul} and {Sironi}}(2020)}]{babul_sironi_20}
\bibinfo{author}{\bibfnamefont{A.-N.} \bibnamefont{{Babul}}} \bibnamefont{and}
  \bibinfo{author}{\bibfnamefont{L.}~\bibnamefont{{Sironi}}},
  \bibinfo{journal}{\mnras} \textbf{\bibinfo{volume}{499}},
  \bibinfo{pages}{2884} (\bibinfo{year}{2020}), \eprint{2006.03081}.

\bibitem[{\citenamefont{{Spitkovsky}}(2005)}]{spitkovsky_05}
\bibinfo{author}{\bibfnamefont{A.}~\bibnamefont{{Spitkovsky}}}, in
  \emph{\bibinfo{booktitle}{Astrophysical Sources of High Energy Particles and
  Radiation}}, edited by \bibinfo{editor}{\bibnamefont{{T.~Bulik, B.~Rudak, \&
  G.~Madejski}}} (\bibinfo{year}{2005}), vol. \bibinfo{volume}{801} of
  \emph{\bibinfo{series}{AIP Conf. Ser.}}, p. \bibinfo{pages}{345},
  \eprint{arXiv:astro-ph/0603211}.

\bibitem[{\citenamefont{{Hoshino}}(2001)}]{hoshino_01}
\bibinfo{author}{\bibfnamefont{M.}~\bibnamefont{{Hoshino}}},
  \bibinfo{journal}{Progress of Theoretical Physics Supplement}
  \textbf{\bibinfo{volume}{143}}, \bibinfo{pages}{149} (\bibinfo{year}{2001}).

\bibitem[{\citenamefont{{Drake} et~al.}(1974)\citenamefont{{Drake}, {Kaw},
  {Lee}, {Schmid}, {Liu}, and {Rosenbluth}}}]{drake_74}
\bibinfo{author}{\bibfnamefont{J.~F.} \bibnamefont{{Drake}}},
  \bibinfo{author}{\bibfnamefont{P.~K.} \bibnamefont{{Kaw}}},
  \bibinfo{author}{\bibfnamefont{Y.~C.} \bibnamefont{{Lee}}},
  \bibinfo{author}{\bibfnamefont{G.}~\bibnamefont{{Schmid}}},
  \bibinfo{author}{\bibfnamefont{C.~S.} \bibnamefont{{Liu}}}, \bibnamefont{and}
  \bibinfo{author}{\bibfnamefont{M.~N.} \bibnamefont{{Rosenbluth}}},
  \bibinfo{journal}{Physics of Fluids} \textbf{\bibinfo{volume}{17}},
  \bibinfo{pages}{778} (\bibinfo{year}{1974}).

\bibitem[{\citenamefont{{Sobacchi} et~al.}(2021)\citenamefont{{Sobacchi},
  {Lyubarsky}, {Beloborodov}, and {Sironi}}}]{sobacchi_21}
\bibinfo{author}{\bibfnamefont{E.}~\bibnamefont{{Sobacchi}}},
  \bibinfo{author}{\bibfnamefont{Y.}~\bibnamefont{{Lyubarsky}}},
  \bibinfo{author}{\bibfnamefont{A.~M.} \bibnamefont{{Beloborodov}}},
  \bibnamefont{and} \bibinfo{author}{\bibfnamefont{L.}~\bibnamefont{{Sironi}}},
  \bibinfo{journal}{\mnras} \textbf{\bibinfo{volume}{500}},
  \bibinfo{pages}{272} (\bibinfo{year}{2021}), \eprint{2010.08282}.

\bibitem[{Note2()}]{Note2}
Note2, \bibinfo{note}{we remark, however, that in \protect \citet {iwamoto_18}
  the O-mode emission was attributed to gyro-phase bunching in Weibel-generated
  fields, which cannot operate in the high-$\sigma $ shocks presented in this
  Letter.}

\bibitem[{\citenamefont{{P{\'e}tri} and
  {Lyubarsky}}(2007)}]{petri_lyubarsky_07}
\bibinfo{author}{\bibfnamefont{J.}~\bibnamefont{{P{\'e}tri}}} \bibnamefont{and}
  \bibinfo{author}{\bibfnamefont{Y.}~\bibnamefont{{Lyubarsky}}},
  \bibinfo{journal}{\aap} \textbf{\bibinfo{volume}{473}}, \bibinfo{pages}{683}
  (\bibinfo{year}{2007}), \eprint{0707.1782}.

\bibitem[{Note3()}]{Note3}
Note3, \bibinfo{note}{even though we focus on the leading soliton, the features
  described below are observed throughout the density cavity until the second
  soliton.}

\bibitem[{\citenamefont{{Rybicki} and {Lightman}}(1979)}]{rybicki_lightman_79}
\bibinfo{author}{\bibfnamefont{G.~B.} \bibnamefont{{Rybicki}}}
  \bibnamefont{and} \bibinfo{author}{\bibfnamefont{A.~D.}
  \bibnamefont{{Lightman}}}, \emph{\bibinfo{title}{{Radiative Processes in
  Astrophysics}}} (\bibinfo{publisher}{John Wiley \& Sons, Inc.},
  \bibinfo{year}{1979}).

\bibitem[{Note4()}]{Note4}
Note4, \bibinfo{note}{detailed calculations of the observed polarization degree
  from FRB-producing shocks, accounting for both shock curvature and
  propagation effects, will be presented elsewhere.}

\end{thebibliography}

\end{document}